\documentclass[preprint,showpacs,preprintnumbers,amsmath,amssymb]{revtex4}

\usepackage{graphicx}
\usepackage{dcolumn}
\usepackage{bm}
\usepackage{textcomp}
\usepackage{ulem}

\usepackage{amstext}

\begin{document}

\preprint{APS/123-QED}

\title{Low-Voltage Nano-Domain Writing in He-Implanted Lithium Niobate Crystals}

\author{M. Lilienblum$^1$}
\author{A. Ofan$^2$}
\author{\'{A}. Hoffmann$^1$}
\author{O. Gaathon$^2$}
\author{L. Vanamurthy$^3$}
\author{S. Bakhru$^3$}
\author{H. Bakhru$^3$}
\author{R. M. Osgood, Jr.$^2$}
\author{E. Soergel$^1$}
\email{soergel@uni-bonn.de}

\affiliation{$^1$Institute of Physics, University of Bonn, Wegelerstra\ss e 8, 53115 Bonn, Germany}
\affiliation{$^2$Center for Integrated Science and Technology, Columbia University, New York NY, 10027 USA}
\affiliation{$^3$College of Nanoscale Science and Engineering, State University of New York at Albany, Albany NY, 12222 USA}


\begin{abstract}
A scanning force microscope tip is used to write ferroelectric domains in He-implanted single-crystal lithium niobate and subsequently probe them by piezoresponse force microscopy. Investigation of cross-sections of the samples showed that the buried implanted layer, $\sim 1$\,\textmu m below the surface, is non-ferroelectric and can thus act as a barrier to domain growth. This barrier enabled stable surface domains of $< 1$\,\textmu m size to be written in 500\,\textmu m-thick crystal substrates with voltage pulses of only  10\,V applied to the tip.

\end{abstract}


\maketitle


Nanostructured domain patterns in lithium niobate (LiNbO$_3$) and lithium tantalate (LiTaO$_3$) single crystals have two important applications. In the first, ferroelectric domain patterns are used for high-density data storage. For this case, outstanding results have been obtained by poling and detection of nm-sized domains by using scanning probe microscopy on samples of 70--150\,nm thickness~\cite{Cho02}. The second application uses domain patterns for photonic device structures such as integrated optical circuits~\cite{Soh08} or photonic crystals~\cite{Rou05}. Whereas data storage requires the smallest possible domain size, ferroelectric domain structures for optical purposes need to have sizes comparable to the operating wavelength, i.\,e., $\sim 400$ to 1500\,nm. Domain spreading when poling across standard thickness wafers makes achieving the high resolution needed for certain applications, e.\,g. photonic crystals at visible wavelengths, a challenge. Several publications have reported on the domain writing in  \textmu m's thick (-thin) LiNbO$_3$ crystals by applying moderate voltages ($<100\,$V) to the tip of a scanning force microscope (SFM)~\cite{Xue03,Rod05,Kan07}. However, the fabrication as well as the handling of these, very thin crystals can be problematical in many applications. Other publications have shown that $<1\,$\textmu m-sized domains can be written into 500\,\textmu m thick single crystals, also, via SFM writing; for such thick wafers, poling requires applying very high voltages ($> 1000$\,V) to the tip~\cite{Ros03,Agr04}. In addition, the quality of the domains, including their shape and depth profile, as well as their reproducibility has not generally been sufficient for optical applications. Similar difficulties occurred when fabricating large-area patterns of sub-\textmu m-sized domains in thick crystals by interference photolithography and subsequent electric-field poling~\cite{Gri05}.

Here we present results on domain patterning in a micrometer-scale surface layer of bulk single-crystal LiNbO$_3$, corresponding to the thickness of most guided-wave or photonic-crystal layers~\cite{Tan02,Rou05,Che08}. Domain patterning was carried out after first implanting the wafer with hundreds keV He ions and then applying low- to moderate-voltage pulses to the tip of a SFM. The written domains were probed and imaged using piezoresponse force microscopy (PFM). The process was optimized via a systematic variation of He-implantation energy and flux in combination with examining changes in the writing parameters, i.\,e., the amplitude and duration of the applied voltage pulse.

In our experiments we used congruent samples of 500\,\textmu m thick, single-domain LiNbO$_3$ crystals (Crystal Technology). The samples were He-implanted through either the $-z$ or the $+z$ face with a fluence of~$F=10^{14}$ to $10^{16}$\,$\rm cm^{-2}$. The energy~$E$ of the ions was chosen to give an end of range (EOR) compatible with the depth resolution of PFM~\cite{Joh09}, thus the samples were implanted at 130 and 350\,keV, corresponding to EOR of 0.5 and 1\,\textmu m, respectively. A more detailed description of the sample preparation can be found elsewhere~\cite{Rad99}.

The experiments were carried out with a commercial scanning probe microscope (Solaris, NT-MDT) using conductive diamond-coated probes (DCP11, NT-MDT). Using a custom-designed script we could automatically write a grid of domains varying poling voltage $V$ and duration $t$ (termed ''poling map'' below), for a range of $V$ from $-100$ to $+100$\,V and $t$ from $1$\,ms to $100$\,s. For recording the generated domain patterns via PFM we applied an alternating voltage $V_{\rm ac} = 5\,\rm V_{rms}$ of frequency $f \approx 30\,$kHz to the tip. A more detailed description of PFM is contained in references~\cite{Alexe,Jun06}.

\begin{figure}[hhh]
\includegraphics{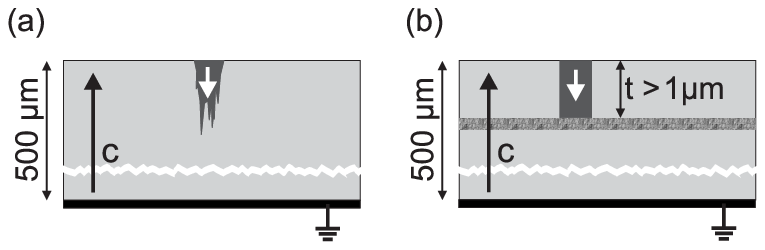}
\caption{\label{fig:lil01} Comparison between an untreated
(a) and an He-implanted (b) LiNbO$_3$ sample with respect to
specific surface domain characteristics.}
\end{figure}

The effect of He-implantation on the surface domain properties is schematically depicted in Fig.~\ref{fig:lil01} where (a) shows the situation for a virgin, untreated sample and (b) for a He-implanted sample. In case of the unimplanted sample, the surface domain forms a direct head-to-head (or tail-to-tail) configuration. This domain configuration is energetically unfavorable~\cite{Lu06}, as a result writing of surface domains in untreated LiNbO$_3$ with the tip yields shallow spiked-shaped domains. In case of the He-implanted sample, however, the He-stopping layer acts as a screening layer between the surface domain and the original bulk, allowing for easy domain writing with the tip. We attribute this result to the fact that the ferroelectricity is strongly reduced in the He-stopping layer; see below. In addition, He-implantation provides a higher concentration of domain termination points, thereby stabilizing surface domains. Note that although the applied electric field within the crystal is the same for both samples shown in Fig.~\ref{fig:lil01}, \textit{stable}, i.\,e. permanent, domains of predetermined size could only be formed in the He-implanted sample.

\begin{figure}[hhh]
\includegraphics{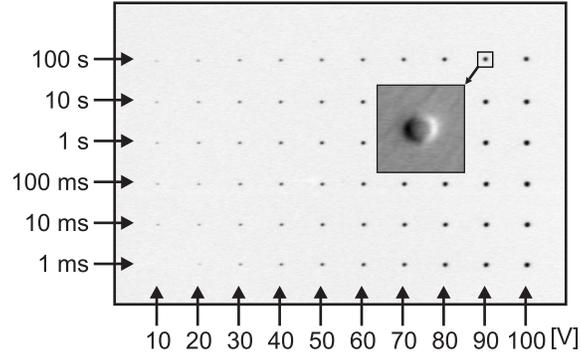}
\caption{\label{fig:lil02}  PFM image (vertical image) of a poling map on a $+z$-faced He-irradiated sample  ($F=10^{14}$\,cm$^{-2}$, $E = 130$\,keV).
The image size is $\rm 33 \times 22\,$\textmu m$^2$. The inset shows a zoom on one domain as lateral PFM image ($\rm 1.2 \times 1.2\,$\textmu m$^2$). The contrast at the domain boundaries evidences the written structures to be ferroelectric domains.}
\end{figure}

A typical poling map used to examine the variation of poling properties of the He-implanted samples with voltage and time is displayed in Fig.~\ref{fig:lil02}, for the case of the He-implantation through the $+z$-face of the crystal. The amplitude of the voltage pulses applied via the tip was varied from 10\,V to 100\,V and their duration was altered over 5\,orders of magnitude (from 1\,ms to 100\,s). Domain imaging was performed by recording the vertical and the lateral PFM signal simultaneously~\cite{Jun09}; for the top overview we chose the vertical image while for the inset a lateral image was used. The contrast at the domain boundary in the lateral image (left: black, right: white) shows clearly that the written structures are ferroelectric domains. Note that all domains in this map were stable, i.\,e., the poling map could be imaged for greater than 1\,month after writing.

The poling map shows clearly that in implanted samples the domains can be generated over the full range of the parameters used here, with the exception of the operating point of 10\,V at 1\,ms. These results are similar to the  writing parameters that have been observed using only  $< 1\,$\textmu m thick stoichiometric LiNbO$_3$ samples~\cite{Rod05}. Note that the coercive field of stoichiometric LiNbO$_3$ is smaller by one order of magnitude when compared to that of the congruent crystals used here~\cite{Wen04}. Observations by other groups using unimplanted samples of 5\,\textmu m and 500\,\textmu m thicknesses showed that much higher voltages and longer writing times than used here are needed for reproducible domain generation~\cite{Ter03,Ros03}.
%
Poling maps written for samples implanted through their -z-face resulted in hampered and less reproducible domains compared to those written on +z-face implanted samples. This difference might be attributed to the unfavorable combination of the inhomogeneous electric field emerging from the tip and preferential starting for domain growth at the +z-face.

In our experiments, the size of the domains was measured as a function of writing voltage~$V$ and pulse width~$t$. These measurements are shown in Fig.~\ref{fig:lil02}; note that the radius of the domains increased linearly with the voltage applied to the tip, a behavior reported in other publications~\cite{Xue03,Rod05}. Note that the slope of this linear dependence was found to vary strongly with the specific sample and probe. Interestingly, although domain size has been commonly reported to depend on the pulse duration~\cite{Xue03,Rod05,Jen09}, our measurements, repeated over many experiments, showed that there was no such pulse-duration dependence. One tentative explanation is related to the limited mobility of space charge in the implantation layer, however more extensive studies must be done to understand this phenomenon. In addition, poling occurred readily within the volume, in which the electric field from the tip exceeded the coercive field since no sideways domain growth was observed upon longer application of the writing voltage. We attribute this behavior to the high density of defects allowing for easy and smooth poling and therefore eliminating the otherwise crucial dependence on the pulse duration. Note that prior experiments with through-wafer He-irradiation of 5~mol\% Mg doped congruent LiNbO$_3$ samples ($E = 41$\,MeV, $F = 10^{12}\,\rm cm^{-2}$) showed that ion irradiation decreased the coercive field by 20\% for tip-induced poling ~\cite{Jen08}. This field decrease was attributed to ion-induced defects, which augmented the density of poling nuclei. Finally, our measurements also showed that the maximum domain size depended on the distance between adjacent domains, indicating an electrostatic repulsion that limits further domain growth~\cite{Ofa09}.


Domain geometry plays an important role in applications of local poling. As seen in Fig.~\ref{fig:lil02}, top-view images of the shape of the domains showed approximately circular images. Although in LiNbO$_3$ the crystals' symmetry strongly favors hexagonal domain growth, it is known that in case of very small characteristic diameters other factors can effect the domain geometry~\cite{Ter03,Rod05}. Moreover simple resolution issues of PFM imaging impede precise mapping of the domains geometry; thus the lateral resolution with the tips used here (radius 50--70\,nm) is limited to $\sim$~75\,nm~\cite{Jun08}. In addition, the depth of the domains can be estimated from the PFM contrast when compared to images recorded on a reference sample with bulk domains (PPLN)~\cite{Joh09}. From the contrast observed here, we conclude that the domains grow through the full over-layer and up to the stopping layer. Note that the EOR for the sample in Fig.~\ref{fig:lil02} is 500\,nm, which limits the maximum depth of the written domains.

\begin{figure}[hhh]
\includegraphics{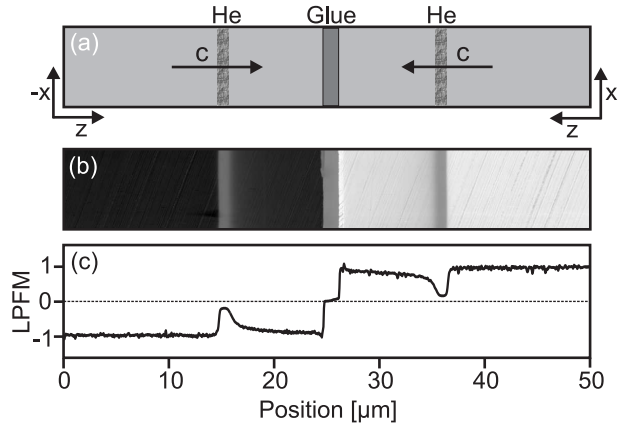}
\caption{\label{fig:lil03}  Schematics of the cross-section
sample (a), a lateral piezoresponse force microscopy
(LPFM) image of the cross-section (b), and a scanline (c)
obtained for a sample implanted by $F = 5\times10^{16}$\,He\,cm$^{-2}$ at
$E = 3.8$\,MeV. The grey areas in (b) indicate no LPFM signal  corresponding to zero in (c).}
\end{figure}

In order to further examine our conclusions on the role of the He-stopping layer we investigated its ferroelectricity with the help of cross-section samples. A schematic of such a sample is seen in Fig.~\ref{fig:lil03}(a). In particular two samples were glued together and optically polished on the side face to allow for PFM investigations. For this cross-sectional measurement a deeper, 10\,\textmu m implantation depth (3.8\,MeV) was used to enable a better distinction between different regions of the sample. The observed contrast in the lateral PFM (LPFM) image can be explained by in-plane deformation of the sample surface owing to the in-plane electric field from the tip together with the $d_{\rm 33}$ piezoelectric tensor element~\cite{Jun09}. This contrast enabled us to make spatially resolved images of the magnitude of  $d_{\rm 33}$, which remained after He-implantation. The overview of the section together with a corresponding scanline is shown in Fig.~\ref{fig:lil03}(b) and (c), respectively.

In order to provide comparative measurement of the piezoelectricity, it is necessary to first make measurements of the PFM response in regions that were not exposed to implantation beam. Thus full piezoresponse can be probed in regions deeper than the implantation stopping range, i.\,e., regions 0--10\,\textmu m and 40--50\,\textmu m in Fig.~\ref{fig:lil03}. Owing to the opposed direction of the $\vec c$-axis in both parts of the sample those regions are seen as black ($-1$) or white ($+1$). Note also that the region below the surface also exhibited nearly the same piezoelectricity, a result in accord with other crystal and electrical measurements on implanted LiNbO$_3$. As expected the non-piezoelectric glue layer shows no LPFM response, i.\,e., a medium grey level ($0$). Based on these considerations, it is possible to determine the degree of ferroelectric degradation within the He-implanted region on the supposition that a decreased piezoresponse is accompanied by a reduced ferroelectricity.

Obviously from the scanline in Fig.~\ref{fig:lil03}(c) both He-stopping regions show a strongly reduced piezoresponse (only 15\% of the full signal), in agreement with the conclusion that this layer has reduced ferroelectricity and can thus act as a screening layer for head-to-head domains (Fig.~\ref{fig:lil01}(b)). The scanline shows clearly the spatial degradation of the ferroelectricity of the sample due to He-implantation. At the very surface, directly at the intersection with the glue, nearly full piezoresponse is observed. Towards the He-stopping layer, the LPFM signal decreases according to the well-known ion distribution for He-implantation~\cite{Zie06}, and reaches its minimum at the EOR. As expected the edge towards the non-irradiated material of the He-stopping layer is noticeably sharper (Figs~\ref{fig:lil03}(b) and (c)).

In conclusion, we have succeeded in controllably and permanently writing submicrometer-scale ferroelectric domains by means of scanning force microscopy into a thick lithium niobate single crystal by He-imlanting a micrometer-deep layer prior to poling. The domains are stable, and readily written with low poling voltage and short pulse duration. The low poling voltage is attributed
to the role of the He-stopping layer acting as an insulating layer for the head-to-head domain configuration; this explanation is in accord with our measurements of the ferroelectric response of cross-section samples with high spatial resolution.


\begin{acknowledgments}
This work (RO, AO, OG, LV, SB, HB) was supported by the NSF (DMR-08-0668206).
Financial support from the Deutsche Telekom AG  (ES, ML, \'{A}H) is gratefully acknowledged.
\end{acknowledgments}


\clearpage


\clearpage

\begin{center}
\bf Figure 1
\end{center}

\clearpage

\begin{center}
\bf Figure 2
\end{center}

\clearpage

\begin{center}
\bf Figure 3
\end{center}

\end{document}